\newif\ifIEEE
\IEEEtrue
\ifIEEE
  \documentclass[journal]{IEEEtran}

\else
  \documentclass[12pt]{article}
  \usepackage{amsthm}

\fi
\usepackage{amsfonts,bm}
\usepackage{ifthen}
\usepackage{color}

\renewcommand{\mathbf}[1]{{\bm{#1}}}     
\newtheorem{theorem}{\indent Theorem}[section]
\newtheorem{proposition}[theorem]{\indent Proposition}
\newtheorem{lemma}[theorem]{\indent Lemma}
\ifIEEE
  \newcommand{\qed}{\hspace*{\fill}%
  \vbox{\hrule\hbox{\vrule\squarebox{.667em}\vrule}\hrule}\smallskip}
  \def\squarebox#1{\hbox to #1{\hfill\vbox to #1{\vfill}}}
  \newcommand{\Proof}[1]{%
                  \textit{Proof\ifthenelse{\equal{#1}{}}{}{ #1}:}}
  \newcommand{\Endspace}{\vspace{0ex}}   
\else
  \theoremstyle{definition}              
  \newcommand{\Proof}[1]{%
                  \textbf{Proof\ifthenelse{\equal{#1}{}}{}{ #1}.}}
  \newcommand{\Endspace}{}
\fi
\newtheorem{example}{\indent Example}[section]
\newtheorem{remark}{\indent Remark}[section]

\newcommand{\GF}{{\mathrm{GF}}}
\newcommand{\Zring}{{\mathbb{Z}}}
\newcommand{\bldzero}{{\mathbf{0}}}

\newcommand{\bldbeta}{{\textrm{\boldmath{$\beta$}}}}
\newcommand{\bldgamma}{{\textrm{\boldmath{$\gamma$}}}}
\newcommand{\bldxi}{{\textrm{\boldmath{$\xi$}}}}
\newcommand{\bldc}{{\mathbf{c}}}
\newcommand{\blde}{{\mathbf{e}}}
\newcommand{\bldu}{{\mathbf{u}}}
\newcommand{\bldv}{{\mathbf{v}}}
\newcommand{\bldy}{{\mathbf{y}}}
\newcommand{\code}{{\mathcal{C}}}
\newcommand{\Set}{{\mathcal{S}}}
\newcommand{\Subset}{{\mathcal{T}}}
\newcommand{\volume}{{\mathcal{V}}}
\newcommand{\messages}{{\mathcal{M}}}
\newcommand{\encoder}{{\mathcal{E}}}
\newcommand{\decoder}{{\mathcal{D}}}
\newcommand{\RS}{{\mathrm{RS}}}
\newcommand{\Range}{{\mathcal{R}}}

\newcommand{\Title}{List Decoding of Burst Errors}

\newcommand{\Namea}{Ron M. Roth}
\newcommand{\Thnxa}{\Namea\ is with the
                    Computer Science Department,
                    Technion,
                    Haifa 32000, Israel.
                    This work was done in part while visiting
                    Hewlett--Packard Laboratories,
                    1501 Page Mill Road,
                    Palo Alto, CA 94304, USA.
                    \par Email:
                    ronny@cs.technion.ac.il}
\newcommand{\Nameb}{Pascal O. Vontobel}
\newcommand{\Thnxb}{\Nameb\ is with
                    Hewlett--Packard Laboratories,
                    1501 Page Mill Road,
                    Palo Alto, CA 94304, USA.
                    \par Email:
                    pascal.vontobel@ieee.org}
\newcommand{\Version}{Submitted to IEEE Transactions on Information Theory, 
                      August 19, 2008. The material in this paper was
                      presented in part at the IEEE International Symposium on
                      Information Theory, Toronto, Canada, July 2008.}

\begin{document}
\ifIEEE
  \title{\Title}
  \author{\thanks{\Version}
          \Namea\thanks{\Thnxa}
          \quad\quad
          \Nameb\thanks{\Thnxb}
  }
  \maketitle
\else
  \title{\textbf{\Title}\footnotetext{\Version}}
  \author{\textsc{\Namea}\thanks{\Thnxa}
  \and
          \textsc{\Nameb}\thanks{\Thnxb}
  }
  \date{}
  \maketitle
\fi

\begin{abstract}
A generalization of the Reiger bound is presented for
the list decoding of burst errors. It is then shown that
Reed--Solomon codes attain this bound.
\vspace{2ex}

\end{abstract}

\section{Introduction}

Many interesting data transmission and storage systems can be modeled as
channels that introduce burst errors. Assuming a list decoder at the
receiver side, we study requirements that a code must satisfy
in order to be
suitable for data transmission over such channels, in particular, we
investigate lower bounds on the code redundancy. As we will see, the
resulting bounds depend on the structure of the code, i.e., we obtain
different lower bounds for linear codes
and group codes on
the one hand, and for unstructured codes on the other hand.
These bounds can be seen as generalizations of the
classical Reiger bound~\cite{Lin:Costello:83:1, Peterson:Weldon:72:1}.
Finally, we show that Reed--Solomon codes achieve the above-mentioned
redundancy lower bound for linear codes.
For proving this latter result, we
will derive a generalization of the known formula for the resultant
of two polynomials, to a larger number of polynomials that have
a certain structure.

We start by presenting several definitions that will be used
throughout this work.
Let $F$ be an alphabet of size $q \ge 2$ and assume
hereafter without loss of generality that $F$ is a finite Abelian group.
The set of words of length $n$ over $F$ is denoted by $F^n$
(which is a group under the operation of component-by-component
addition of elements of $F$).

We say that a word $\blde \in F^n$
is a \emph{$\tau$-burst} if either $\blde = \bldzero$
(the all-zero word) or the indexes $i$ and $j$ of the first
and last nonzero entries in $\blde$ satisfy $j - i < \tau$.

Let $\code$ be a code of length $n$ over $F$.
A \emph{decoder} for $\code$ is
a mapping $\decoder : F^n \rightarrow 2^\code$,
where $2^\code$ denotes the power set of $\code$.
The \emph{list size} of a decoder $\decoder$ is the largest size
of $\decoder(\bldy)$ over all $\bldy \in F^n$.

We say that $\decoder$ \emph{detects} any single $\tau$-burst error
if for every codeword $\bldc \in \code$ and every
$\tau$-burst $\blde \in F^n$,
\[
\decoder(\bldc + \blde) =
\left\{
\begin{array}{ccl}
\{ \bldc \} && \textrm{if $\blde = \bldzero$} \\
\emptyset   && \textrm{otherwise}
\end{array}
\right.
\; .
\]
Such a decoder for $\code$ exists if and only if
for any two distinct codewords $\bldc_1, \bldc_2 \in \code$,
the difference $\bldc_1 - \bldc_2$ is not a $\tau$-burst.

We say that $\decoder$ \emph{corrects}
any single $\tau$-burst error if for
every codeword $\bldc \in \code$ and every $\tau$-burst $\blde \in F^n$,
\[
\bldc \in \decoder(\bldc + \blde) \; .
\]

An \emph{$(\ell,\tau)$-burst list decoder} for $\code$
is a decoder for $\code$ of list size at most $\ell$
that corrects any single $\tau$-burst error.
Such a decoder exists if and only if
there are no $\ell{+}1$ distinct pairs
\[
(\bldc_0,\blde_0),
(\bldc_1,\blde_1), \ldots,
(\bldc_\ell,\blde_\ell) \; ,
\]
where each $\bldc_i$ is a codeword,
each $\blde_i$ is a $\tau$-burst, and
\[
\bldc_0 + \blde_0 = 
\bldc_1 + \blde_1 = \cdots =
\bldc_\ell + \blde_\ell \; .
\]

For the case $\ell = 1$ (conventional single $\tau$-burst decoding),
we have the well-known Reiger bound, which states that
if a code $\code$ has a $(1,\tau)$-burst list decoder then
the redundancy of $\code$,
\[
r = n - \log_q |\code| \; ,
\]
is at least $2 \tau$ (the bound is usually stated for
linear codes---see for example~\cite[p.~258]{Lin:Costello:83:1}
or~\cite[p.~110]{Peterson:Weldon:72:1}---although it holds
for nonlinear codes as well).

The Reiger bound holds even under the restriction that the burst errors
are \emph{phased}~\cite[p.~272]{Lin:Costello:83:1},
namely, the support of
the $\tau$-burst error is contained in one of the following sets
$J_i$ (assuming that entry indexes start at $0$):
\begin{equation}
\label{eq:Ji}
J_i = \left\{ j \;:\; i \tau \le j < (i{+}1) \tau \right\} \; ,
\quad
0 \le i < n/\tau \; .
\end{equation}
When non-overlapping $\tau$-blocks over $F$ are regarded as symbols of
the alphabet $F^\tau$, a phased $\tau$-burst error becomes
a single symbol (random) error over $F^\tau$.

When $F$ is a field, then Reed--Solomon codes over $F$ attain
the Reiger bound and, in fact, they are optimal also for
the deterministic correction of multiple burst errors
(for probabilistic correction, see~\cite{Krachkovsky:03:1}).

Building upon a result by
Parvaresh and Vardy~\cite{Parvaresh:Vardy:05:1},
Guruswami and Rudra presented in~\cite{Guruswami:Rudra:06:1}
a construction of codes that have a polynomial-time list decoder
that corrects any pattern of up
to $r (1-\varepsilon)$ errors, where $r$ is the code redundancy and
$\varepsilon$ is any fixed small positive real.
The Guruswami--Rudra scheme is, in fact,
a list decoder for Reed--Solomon codes that corrects multiple
\emph{phased} burst errors.

In this work, we consider the problem of list decoding
of single burst errors that are not necessarily phased.
In Section~\ref{sec:Reigerbound}, we present lower bounds
on the redundancy of codes that have $(\ell,\tau)$-burst list decoders.
In most cases, we will assume that the code also has a decoder
that detects any single $\tau$-burst error.
In Sections~\ref{sec:tools}--\ref{sec:RSbound},
we show that Reed--Solomon codes attain the respective lower bound
for linear codes.

\begin{remark}
In practice, the code $\code$ serves as the set of images of
an \emph{encoding mapping} $\encoder : \messages \rightarrow \code$,
where $\messages$ is the set of messages to be transmitted
through the (noisy) channel.
In the context of list decoding, the mapping $\encoder$ does not have
to be lossless (i.e., one-to-one), but then, in determining
the list size of a decoder $\decoder$, we need to count
each codeword $\bldc$ in $\decoder(\bldy)$
a number of times which equals the number of pre-images of $\bldc$
in $\messages$ (equivalently, the list size is
the largest number of distinct \emph{messages} that are returned by
the decoder).
However, when using a many-to-one encoder, the decoding can be
ambiguous even when no errors have occurred.
Such a feature is undesirable in virtually all practical
applications: if ambiguity is to be allowed (through the decoding into
a list of size greater than $1$), then it should be limited only
to cases where errors have occurred---as the probability of
that to happen is presumed to be small (yet not negligible).
Therefore, our definition of the list size of $\decoder$
assumes that the encoding is lossless,
thereby allowing us to regard codewords as messages.
And, as said earlier, we will also want the decoder to be able to
tell whether a burst error has occurred.\qed
\end{remark}

\begin{remark}
Since we focus in this paper on the case of a single burst error,
any $(\ell,\tau)$-burst list decoder can be implemented
by enumerating over the location of the first nonzero entry in
the burst error, thereby effectively transforming the burst error
into a burst \emph{erasure}. Now, in the case of linear codes,
erasure decoding amounts to computing a syndrome and solving linear
equations and, so, erasures can be decoded in polynomial time.
Hence, $(\ell,\tau)$-burst list decoders for linear codes
always have a polynomial-time implementation
(although for some linear codes we may get faster implementations
by taking advantage of the specific structure of the code).\qed
\end{remark}

\section{Generalized Reiger Bound}
\label{sec:Reigerbound}

Most of the section will be devoted to generalizing the classical
Reiger bound to our list-decoding setup.
Interestingly, as we have already mentioned,
the resulting lower bounds depend on the structure of the code.
We emphasize
that these differences in lower bounds are not
spurious: we will show (by example)
that there are indeed
unstructured codes whose redundancy is lower than the redundancy
that is required for group or linear codes.

For completeness reasons, we start this section by presenting
a generalization of the classical sphere-packing bound
to our list-decoding setup.
However, unless the codes are long, namely have a block length of at
least $\ell \cdot q^{\tau/\ell}$, this generalized sphere-packing bound
will not be better than the generalized Reiger bound.

\subsection{Sphere-Packing Type Bound}

Given an alphabet $F$ of size $q$,
denote by $\volume_q(n,\tau)$ the number of $\tau$-bursts in $F^n$;
for $0 \le \tau \le n$, this number is given by
\[
\volume_q(n,\tau) =
1 +
(q{-}1)n +
(q{-}1)^2 \sum_{i=0}^{\tau-2} (n{-}i{-}1) q^i \; .
\]
The following sphere-packing type bound for burst list decoding
is proved very similarly to its symbol-error counterpart
in~\cite{Elias:91:1}.

\begin{theorem}
\label{thm:Spherepackingbound}
Let $\code$ be a code of length $n$ over
an alphabet of size $q \ge 2$ and
let $\tau$ and $\ell$ be positive integers.
Then $\code$ has an $(\ell,\tau)$-burst list decoder
only if the redundancy $r$ of $\code$ satisfies
\[
r \ge \log_q \left( \frac{\volume_q(n,\tau)}{\ell} \right) \; .
\]%
\Endspace
\end{theorem}

For $n > 1$, the lower bound in Theorem~\ref{thm:Spherepackingbound}
is smaller than $\tau + \log_q (n/\ell)$.
In this section, we obtain Reiger-type bounds, which turn out to be
better for lengths $n$ that are smaller than $\ell \cdot q^{\tau/\ell}$.

\subsection{Generalized Reiger Bound for Group Codes}
\label{sec:gen:reiger:bound:group:codes:1}

A code $\code$ of length $n$ over (a finite Abelian group) $F$
is called a \emph{group code} over $F$
if it is a subgroup of the group $F^n$.
In particular, if $F$ is a field, then every linear code over $F$ is
a group code over $F$.

For group codes, the conditions for the existence of decoders that
detect or correct any single $\tau$-burst are simplified.
Specifically,
a group code $\code$ has a decoder that detects any single
$\tau$-burst if and only if the all-zero codeword is
the only $\tau$-burst in $\code$.
And such a code has an $(\ell,\tau)$-burst list decoder if and only if
no $\ell{+}1$ distinct $\tau$-bursts belong to
the same coset of $\code$ within $F^n$.
In particular, if $\code$ is a linear code over a field $F$,
then these $\tau$-bursts cannot have the same syndrome
(with respect to any parity-check matrix of $\code$).

The following theorem is a generalization of the Reiger bound
to burst list decoders for group codes.

\begin{theorem}
\label{thm:Reigerbound}
Let $\code$ be a group code of length $n$ over $F$
and let $\tau$ and $\ell$ be positive integers that satisfy
the following three conditions:
\begin{enumerate}
\item
\label{item:bound1}
$(\ell{+}1) \tau \le n$.
\item
\label{item:bound2}
There is a decoder for $\code$ that detects
any single $\tau$-burst error.
\item
\label{item:bound3}
There is an $(\ell,\tau)$-burst list decoder for $\code$.
\end{enumerate}
Then the redundancy $r$ of $\code$ satisfies
\[
r \ge \Bigl( 1 + \frac{1}{\ell} \Bigr) \tau \; .
\]%
\Endspace
\end{theorem}

\Proof{}
Our proof strategy will be to show that if $r$ is not large enough, 
then we can exhibit $\ell{+}1$ distinct pairs $(\bldc_i,\blde_i)$
of codewords $\bldc_i$ and $\tau$-bursts $\blde_i$
that add up to the same word.

Writing $q = |F|$, we therefore suppose that
$r < (\ell{+}1)\tau/\ell$, or, equivalently,
\begin{equation}
\label{eq:contradicting}
\left( \frac{q^n}{|\code|} \right)^\ell < q^{(\ell+1)\tau} \; .
\end{equation}
Let $J_0, J_1, \ldots, J_\ell$ be disjoint subsets of integers
where each $J_i$ consists of $\tau$ consecutive elements from
$\{ 0, 1, \ldots, n{-}1 \}$;
condition~\ref{item:bound1} indeed guarantees that such subsets exist.
For $i = 0, 1, \ldots, \ell$, denote by $\Set_i$ the set of all words
in $F^n$
whose support is contained in $J_i$, and define the set $\Set$ by
\begin{eqnarray*}
\lefteqn{
\Set =
\left\{
( \, \bldv_1{-}\bldv_0 \;|\; \bldv_2{-}\bldv_1 \;|\; \ldots \;|\;
\bldv_\ell{-}\bldv_{\ell-1} \, ) \;:
\right.
} \makebox[15ex]{} \\
&&
\left.
\bldv_i \in \Set_i \quad \textrm{for $i = 0, 1, \ldots, \ell$}
\right\}
\; .
\end{eqnarray*}
Note that $\Set$ is a subset of
\[
(F^n)^\ell =
\underbrace{F^n \times F^n \times \cdots \times F^n}_%
                                     {\textrm{\scriptsize $\ell$ times}}
\]
and that
\[
|\Set| = \prod_{i=0}^\ell |\Set_i| = q^{(\ell+1)\tau}
> \left( \frac{q^n}{|\code|} \right)^\ell \; ,
\]
where the inequality follows from~(\ref{eq:contradicting}).
This means that $|\Set|$ is greater than the number of cosets of
the subgroup
$\code^\ell = \code \times \code \times \cdots \times \code$
of $(F^n)^\ell$ under
the component-by-component addition of elements of $F^n$.
By the pigeon-hole principle, there must be two distinct elements
in $\Set$, say
\[
\bldv =
( \, \bldv_1{-}\bldv_0 \;|\; \bldv_2{-}\bldv_1 \;|\; \ldots \;|\;
\bldv_\ell{-}\bldv_{\ell-1} \, )
\; \phantom{,}
\]
and
\[
\bldv' =
( \, \bldv'_1{-}\bldv'_0 \;|\; \bldv'_2{-}\bldv'_1 \;|\; \ldots \;|\;
\bldv'_\ell{-}\bldv'_{\ell-1} \, )
\; ,
\]
which are in the same coset of $\code^\ell$.
Write $\blde_i = \bldv_i - \bldv'_i$ for $i = 0, 1, \ldots, \ell$;
then $\blde_i \in \Set_i$ for all $i$ and
\begin{equation}
\label{eq:badword}
( \, \blde_1{-}\blde_0 \;|\; \blde_2{-}\blde_1 \;|\; \ldots \;|\;
\blde_\ell{-}\blde_{\ell-1} \, ) = \bldv - \bldv' \in \code^\ell \; .
\end{equation}

Next, we claim that $\blde_i \ne \bldzero$ for all $i < \ell$.
Otherwise, since $\bldv \ne \bldv'$,
there had to be an index $i < \ell$ for which
$\blde_i = \bldzero$ yet $\blde_{i+1} \ne \bldzero$.
But then,
\[
\blde_{i+1} - \blde_i  = \blde_{i+1} \in \code \cap \Set_{i+1} \; ,
\]
thereby contradicting condition~\ref{item:bound2},
as $\code$ would have a codeword that is a nonzero $\tau$-burst.
(It can be easily seen that $\blde_\ell$ is nonzero also, but
we will not need this fact in the sequel.)

As our next step, we claim that $\blde_i \ne \blde_j$ for all
$0 \le i < j \le \ell$:
indeed, since $\Set_i \cap \Set_j = \{ \bldzero \}$,
then $\blde_i = \blde_j$ implies
that both $\blde_i$ and $\blde_j$ are zero, which is impossible.

For $i = 0, 1, \ldots, \ell$, define the words
$\bldc_0, \bldc_1, \ldots, \bldc_\ell \in F^n$ iteratively by
$\bldc_0 = \bldzero$ and
\[
\bldc_{i+1} = \bldc_i + \blde_i - \blde_{i+1} \; ,
\quad 0 \le i < \ell \; .
\]
Since $\code$ is a group code, it follows from~(\ref{eq:badword})
that each $\bldc_i$ is in fact a codeword of $\code$.
Thus, we have found $\ell{+}1$ distinct pairs
\[
(\bldc_0,\blde_0),
(\bldc_1,\blde_1), \ldots,
(\bldc_\ell,\blde_\ell) \; ,
\]
where each $\bldc_i$ is a codeword of $\code$,
each $\blde_i$ is a $\tau$-burst, and
\[
\bldc_0 + \blde_0 = 
\bldc_1 + \blde_1 = \cdots =
\bldc_\ell + \blde_\ell \; .
\]
This, in turn, contradicts condition~\ref{item:bound3}. \qed

\begin{remark}
\label{rem:linear}
If $\code$ is a linear code over the field $F = \GF(q)$, then
its redundancy $r$ is always an integer. In this case,
the lower bound of Theorem~\ref{thm:Reigerbound} can be written as
\begin{equation}
\label{eq:Reigerboundlinear}
r \ge \tau + \left\lceil \frac{\tau}{\ell} \right\rceil \; .
\end{equation}
Furthermore, when $\code$ is linear and
$\ell < q$, then condition~\ref{item:bound2}
is actually implied by condition~\ref{item:bound3}.\qed
\end{remark}

Observe that in the proof of Theorem~\ref{thm:Reigerbound},
we did not make any assumptions on the sets
$J_0, J_1, \ldots, J_\ell$, other than satisfying the following
two properties:
(i)~these sets are disjoint, and
(ii)~each $J_i$ consists of $\tau$ consecutive elements from
$\{ 0, 1, \ldots, n{-}1 \}$.
If we now select any particular 
$\ell{+}1$ sets $J_0, J_1, \ldots, J_\ell$
that satisfy these two properties,
then Theorem~\ref{thm:Reigerbound} still holds
even if the burst error is restricted \emph{a priori}
to have support that is contained in one of the sets $J_i$.
In particular, if
the subsets $J_i$ are taken as in~(\ref{eq:Ji}),
then we get that Theorem~\ref{thm:Reigerbound} holds also
for the restricted case of phased burst errors.

\begin{remark}
As pointed our earlier, when we regard nonoverlapping $\tau$-blocks
over $F$ as symbols of the alphabet $F^\tau$,
a phased $\tau$-burst error becomes a single symbol error.
Assuming that $\tau$ divides $n$, the proof of
Theorem~\ref{thm:Reigerbound} then implies that the code $\code$, when
regarded as a code of length $n/\tau$ over $F^\tau$, has a decoder that
detects a single error and a list decoder of size $\ell$ that corrects
a single error, only if the redundancy of $\code$ is at least
$\frac{1}{\tau} \cdot \bigl( 1 + \frac{1}{\ell} \bigr) \tau = 1 +
\frac{1}{\ell}$.
In fact, this is precisely the statement we get when we
plug in $\tau = 1$ in Theorem~\ref{thm:Reigerbound}.\qed
\end{remark}

When there is no such \emph{a priori} restriction on the location of
the burst errors,
then condition~\ref{item:bound1} in Theorem~\ref{thm:Reigerbound} 
can include more pairs $(\ell,\tau)$:
the next theorem is a modification of
Theorem~\ref{thm:Reigerbound} where condition~\ref{item:bound1}
is relaxed from $(\ell{+}1) \tau \le n$ to $2 \tau \le n$
for pairs $(\ell,\tau)$ in which $\ell$ divides $\tau$.

\begin{theorem}
\label{thm:Reigerbound'}
Theorem~\ref{thm:Reigerbound} holds also
when condition~\ref{item:bound1} therein is relaxed to include pairs
$(\ell,\tau)$ such that $\ell \,|\, \tau$ and $2\tau \le n$.
\end{theorem}

\Proof{}
Again, the proof strategy will be to show that if $r$ is not large
enough, then we can exhibit $\ell{+}1$ distinct pairs
of codewords and $\tau$-bursts that add up to the same word.

Writing $q = |F|$ and $b = \tau/\ell$, we therefore 
assume that $r < (\ell{+}1)b$, or, equivalently,
\begin{equation}
\label{eq:contradicting'}
|\code| = q^{n-r} > q^{n - (\ell+1)b} \; .
\end{equation}
Next, we partition $\code$ into $q^{n-2\tau}$ subsets
$\code(\bldv)$, where $\bldv$ ranges over $F^{n-2\tau}$:
each subset $\code(\bldv)$ consists of all codewords of $\code$
whose $(n{-}2\tau)$-suffix equals $\bldv$. Clearly, there is
at least one word $\bldv'$ for which
\[
|\code(\bldv')| \ge
\frac{|\code|}{q^{n-2\tau}} =
\frac{|\code|}{q^{n-2\ell b}} > q^{(\ell-1)b} \; ,
\]
where the strict inequality follows from~(\ref{eq:contradicting'}).
We let $\code'$ denote the set of all $(2\tau)$-prefixes of
the codewords in $\code(\bldv')$; note that $\code'$ is
a code of length $2\tau$ over $F$, and since $\code$
satisfies the conditions of the theorem, then so does $\code'$.

Let $J_0, J_1, \ldots, J_\ell$ be defined by
\[
J_i = \left\{ j \;:\; i b \le j < i b + \tau \right\} \; ,
\quad
0 \le i \le \ell \; .
\]
For every $i < \ell$ we have
$|J_i \cup J_{i+1}| = \tau + b = (\ell{+}1)b$. 
Since the length of $\code'$ is $2\tau = 2 \ell b$ and its size is
greater than $q^{(\ell-1)b}$, we conclude by
the pigeon-hole principle that $\code'$ must contain
two distinct codewords, say $\bldu_i$ and $\bldu'_i$, which agree on
all positions except possibly those that are indexed by
$J_i \cup J_{i+1}$.

\begin{figure*}[hbt]
\[
\newcommand{\B}{\multicolumn{6}{|c|}{}}
\newcommand{\y}[1]{\makebox[4ex]{$\bldy_{#1}$}}
\begin{array}{cc|cccccccccccc|}
\cline{3-14}
\bldc_0 & &
\B & \y{\ell+1} & \y{\ell+2} & \cdots &
\y{2\ell-2} & \y{2\ell-1} &\y{2\ell} \\
\cline{3-14}
\bldc_1 & &
\y{1} & \B & \y{\ell+2} & \y{\ell+3} &\cdots& \y{2\ell-1} & \y{2\ell} \\
\cline{3-14}
\bldc_2 & &
\y{1} & \y{2} & \B & \y{\ell+3} & \y{\ell+4} & \cdots & \y{2\ell} \\
\cline{3-14}
\multicolumn{14}{c}{} \\
\vdots & \multicolumn{1}{c}{} & \vdots & \vdots & & \ddots &
\multicolumn{5}{c}{} & \ddots && \multicolumn{1}{c}{\vdots} \\
\multicolumn{14}{c}{} \\
\cline{3-14}
\bldc_{\ell-1} & &
\y{1} & \y{2} & \cdots & \y{\ell-2} & \y{\ell-1} & \B & \y{2\ell} \\
\cline{3-14}
\bldc_\ell & &
\y{1} & \y{2} & \y{3} & \cdots & \y{\ell-1} & \y\ell & \B \\
\cline{3-14}
\end{array}
\]
\caption{Configuration of
the codewords $\bldc_0, \bldc_1, \ldots, \bldc_\ell$.}
\label{fig:badcodewords}
\end{figure*}

For $i = 0, 1, \ldots, \ell$, define the codewords
$\bldc_0, \bldc_1, \ldots, \bldc_\ell \in \code'$ iteratively by
$\bldc_0 = \bldzero$ and
\[
\bldc_{i+1} = \bldc_i + \bldu_i - \bldu'_i \; ,
\quad 0 \le i < \ell \; .
\]
Thus, for every $i < \ell$,
the codewords $\bldc_i$ and $\bldc_{i+1}$ agree on
all positions except possibly those that are indexed by
$J_i \cup J_{i+1}$.

Let $\bldy \in F^{2\tau}$ be such that it agrees with $\bldc_0$
on its last $\tau \; (= \ell b)$ positions and
with $\bldc_\ell$ on its first
$\tau$ positions. Write
\[
\bldy =
( \, \bldy_1 \;|\; \bldy_2 \;|\; \ldots \;|\; \bldy_{2\ell} \, )
\; ,
\]
where each $\bldy_j$ is a $b$-block over $F$. From the construction of
the codewords $\bldc_i$ we get by a simple backward induction on $i$
that the $(ib)$-prefix of $\bldc_i$ is given by
\[
\newcommand{\y}[1]{\makebox[3ex]{$\bldy_{#1}$}}
( \, \y{1} \;|\; \y{2} \;|\; \ldots \;|\; \y{i} \, ) \; .
\]
Similarly, by a forward induction on $i$ it follows that
the $((\ell{-}i)b)$-suffix of $\bldc_i$ is given by
\[
\newcommand{\y}[1]{\makebox[6ex]{$\bldy_{#1}$}}
( \, \y{\ell+i+1} \;|\; \y{\ell+i+2} \;|\; \ldots \;|\;
\y{2\ell} \, ) \; .
\]
Thus, the configuration of the codewords
$\bldc_0, \bldc_1, \ldots, \bldc_\ell$ is as shown
in Figure~\ref{fig:badcodewords}.

Define $\blde_i = \bldy - \bldc_i$. From Figure~\ref{fig:badcodewords}
we readily see that the support of $\blde_i$ is contained in $J_i$
and, so, $\blde_i$ is a $\tau$-burst. Obviously,
\[
\bldc_0 + \blde_0 = 
\bldc_1 + \blde_1 = \cdots =
\bldc_\ell + \blde_\ell \; (= \bldy) \; ,
\]
which means that we will establish the contradiction once we show that
the codewords $\bldc_0, \bldc_1, \ldots, \bldc_\ell$ are all distinct.
Indeed, suppose that $\bldc_0, \bldc_1, \ldots, \bldc_i$
are distinct yet $\bldc_{i+1} = \bldc_m$ for some $m \le i$.
Since $\bldc_{i+1} - \bldc_i = \bldu_i - \bldu'_i \ne \bldzero$,
we must actually have $m < i$. But then it follows from
Figure~\ref{fig:badcodewords} that the two (distinct) codewords
$\bldc_i$ and $\bldc_m$ would share the $\ell$ blocks
\[
\bldy_1, \bldy_2, \ldots, \bldy_i,
\quad \textrm{and} \quad
\bldy_{\ell+i+1}, \bldy_{\ell+i+2}, \ldots, \bldy_{2\ell}
\]
and, as such, they would differ on at most $\tau$ positions,
thereby contradicting condition~\ref{item:bound2}.\qed

\begin{remark}
One may ask if condition~\ref{item:bound1} in
Theorems~\ref{thm:Reigerbound} and~\ref{thm:Reigerbound'} can be
further relaxed to requiring only that
$2 \tau \le n$ (without restricting
$\tau$ to be an integer multiple of $\ell$).
The code we present in Appendix~\ref{sec:appA} shows that, in general,
Theorems~\ref{thm:Reigerbound} and~\ref{thm:Reigerbound'} no longer
hold
under such a relaxation.\qed
\end{remark}

\subsection{Generalized Reiger Bound for General Codes}

The lower bound on the redundancy in Theorems~\ref{thm:Reigerbound}
and~\ref{thm:Reigerbound'} applies to group codes.
As the next example shows, this bound does not apply to general codes.

\begin{example}
\label{ex:n=4,tau=ell=2}
Let $F$ be an alphabet of size $q \ge 2$ and consider
the code $\code$ of length $4$ and size $2q{-}2$ over $F$
which is defined as the union of the following two sets:
\[
\code_1 =
\Bigl\{ (a \, a \, a \, 0) \;:\; a  \in F \setminus \{ 0 \} \Bigr\}
\; \phantom{.}
\]
and
\[
\code_2 =
\Bigl\{ (0 \, a \, a \, a) \;:\; a  \in F \setminus \{ 0 \} \Bigr\}
\; .
\]
We claim that $\code$ satisfies
conditions~\ref{item:bound2}--\ref{item:bound3} of
Theorem~\ref{thm:Reigerbound}, for $\tau = \ell = 2$.
Indeed, every two distinct codewords $\bldc_1, \bldc_2 \in \code$
either differ on each of their first three positions
(if $\bldc_1, \bldc_2 \in \code_1$),
or on each of their last three positions
(if $\bldc_1, \bldc_2 \in \code_2$),
or on both their first and last positions
(if $\bldc_1 \in \code_1$ and $\bldc_2 \in \code_2$);
in either case, the difference $\bldc_1 - \bldc_2$ is
not a $2$-burst and therefore condition~\ref{item:bound2} is satisfied.

As for condition~\ref{item:bound3},
suppose to the contrary that there exist three distinct codewords
$\bldc_0, \bldc_1, \bldc_2 \in \code$
and respective three $2$-bursts
$\blde_0, \blde_1, \blde_2 \in F^4$ such that
\[
\bldc_0 + \blde_0 = \bldc_1 + \blde_1 = \bldc_2 + \blde_2 \; .
\]
Since $\code$ has been shown to satisfy condition~\ref{item:bound2},
the supports of $\blde_0$, $\blde_1$, and $\blde_2$ have to be
distinct, which means that
$\bldc_0$, $\bldc_1$, and $\bldc_2$ can be assumed to take
the form shown in Figure~\ref{fig:badcodewords},
with $\bldy_0$, $\bldy_1$, $\bldy_2$, and $\bldy_3$ now being
elements of $F$.
In particular, $\bldc_0$ and $\bldc_1$ agree on their last position,
which is possible only if both belong to $\code_1$.
Similarly, $\bldc_1$ and $\bldc_2$ agree on their first position,
implying that both belong to $\code_2$.
Thus, $\bldc_1$ belongs to both $\code_1$ and $\code_2$,
which is a contradiction since these sets are disjoint.

Now, the redundancy of $\code$ equals $4 - \log_q (2q{-}2)$
and, for $q > 2$, this number is smaller than $3$, which is
the lower bound we get for $\tau = \ell = 2$
in Theorem~\ref{thm:Reigerbound'}.\qed
\end{example}

In fact, Example~\ref{ex:n=4,tau=ell=2} attains the lower bound
in the next result (which applies to list size $2$;
we will generalize this bound to larger $\ell$
in Theorem~\ref{thm:Reigerboundeveryell} below).

\begin{figure*}[hbt]
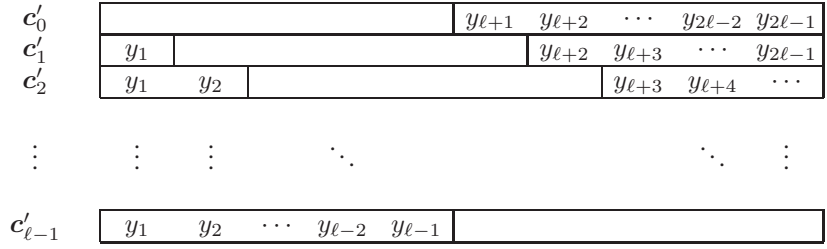

\[
\newcommand{\B}{\multicolumn{5}{|c|}{}}
\newcommand{\y}[1]{\makebox[4ex]{$y_{#1}$}}
\begin{array}{cc|cccccccccc|}
\cline{3-12}
\bldc'_0 & &
\B & \y{\ell+1} & \y{\ell+2} & \cdots &
\y{2\ell-2} & \y{2\ell-1} \\
\cline{3-12}
\bldc'_1 & &
\y{1} & \B & \y{\ell+2} & \y{\ell+3} &\cdots& \y{2\ell-1} \\
\cline{3-12}
\bldc'_2 & &
\y{1} & \y{2} & \B & \y{\ell+3} & \y{\ell+4} & \cdots \\
\cline{3-12}
\multicolumn{12}{c}{} \\
\vdots & \multicolumn{1}{c}{} & \vdots & \vdots & & \ddots &
\multicolumn{4}{c}{} & \ddots & \multicolumn{1}{c}{\vdots} \\
\multicolumn{12}{c}{} \\
\cline{3-12}
\bldc'_{\ell-1} & &
\y{1} & \y{2} & \cdots & \y{\ell-2} & \y{\ell-1} & \B \\
\cline{3-12}
\end{array}
\]
\caption{Configuration of
the words $\bldc'_0, \bldc'_1, \ldots, \bldc'_{\ell-1}$.}
\label{fig:badcodewords'}
\end{figure*}

\begin{proposition}
\label{prop:Reigerboundell=2}
Let $\code$ be a code of length $n$ over an alphabet of size $q \ge 2$
and let $\tau$ be a positive integer that satisfies
the following three conditions:
\begin{enumerate}
\item
$\tau$ is even and $2\tau \le n$.
\item
\label{item:boundell=2}
There is a decoder for $\code$ that detects
any single $\tau$-burst error.
\item
There is a $(2,\tau)$-burst list decoder for $\code$.
\end{enumerate}
Then the redundancy $r$ of $\code$ satisfies
\begin{eqnarray*}
r &\ge& 2\tau - \log_q \left( 2 q^{\tau/2} - 2 \right) \\
  &=&  \left(
         1 + \frac{1}{2}
       \right)
       \tau
       - \log_q 2
       + \log_q \left( \frac{1}{1 - q^{-\tau/2}} \right)
    \; .
\end{eqnarray*}
In particular, $r > \bigl(1 {+} \frac{1}{2} \bigr) \tau - \log_q 2$.
\end{proposition}

\Proof{}
Write $b = \tau/2$,
and suppose to the contrary that 
$r < 2\tau - \log_q \left( 2 q^b - 2 \right)$;
namely,
\begin{equation}
\label{eq:contradictingell=2}
|\code| = q^{n-r} > q^{n-2\tau} \cdot (2q^b - 2) \; .
\end{equation}
Let $\code'$ be the code of length $2\tau$ as defined in
the proof of Theorem~\ref{thm:Reigerbound'};
recall that since $\code$ satisfies the three conditions of
the theorem, then so does $\code'$. From~(\ref{eq:contradictingell=2})
we get that
\[
|\code'| \ge \frac{|\code|}{q^{n-2\tau}} > 2q^b - 2 \; ,
\]
that is,
\begin{equation}
\label{eq:code'}
|\code'| \ge 2q^b - 1 \; .
\end{equation}

Let $\bldc$ be a codeword of $\code$.
We say that a codeword $\bldc' \ne \bldc$ in $\code$ is
a \emph{right} (respectively, \emph{left}) \emph{neighbor} of $\bldc$
if $\bldc$ and $\bldc'$ share the same suffix
(respectively, prefix) of length $b$.
Let $\code'_1$ (respectively, $\code'_2$) be the set of
all codewords of $\code'$ that have no right
(respectively, left) neighbors. Since the $b$-suffixes of
the elements of $\code'_1$ must all be distinct, we must
have $|\code_1'| \le q^b$. From~(\ref{eq:code'})
it follows that the set $\code' \setminus \code'_1$ is nonempty;
hence, there is at least one $b$-block that does not appear
as a $b$-suffix in any element in $\code'_1$. Thus,
$|\code'_1| \le q^b - 1$ and, since the same upper bound applies
to $|\code'_2|$, we get
\[
\left| \code' \setminus (\code'_1 \cup \code'_2) \right|
\ge (2q^b - 1) - 2(q^b - 1) \ge 1 \; .
\]
We conclude that $\code'$ contains a codeword $\bldc_1$ that
has both a right neighbor $\bldc_0$ and a left neighbor $\bldc_2$,
and by condition~\ref{item:boundell=2} these two neighbors
must be distinct. Yet the codewords
$\bldc_0$, $\bldc_1$, and $\bldc_2$ form the violating configuration
of Figure~\ref{fig:badcodewords}, thereby reaching a contradiction.\qed

The next lemma will be used to generalize
Proposition~\ref{prop:Reigerboundell=2} to larger $\ell$.

\begin{lemma}
\label{lem:Reigerboundtau=ell}
Let $\ell$ be an integer greater than $1$ and
let $\code$ be a code of length $2\ell$ over an alphabet of size $q$.
Suppose that $\code$ satisfies
conditions~\ref{item:bound2}--\ref{item:bound3}
in Theorem~\ref{thm:Reigerbound} for $\tau = \ell$. Then
\[
|\code|  < \ell \cdot q^{\ell-1} \; .
\]%
\Endspace
\end{lemma}

\Proof{}
We prove the lemma by induction on $\ell$.
For any integer $\ell > 1$, we denote by $M(\ell)$ the size of
the largest code $\code$ of length $2\ell$
that satisfies the conditions of the lemma.

The induction base ($\ell = 2$) follows by
substituting $\tau = 2$ and $n = 4$ in
Proposition~\ref{prop:Reigerboundell=2}: we get $M(2) \le 2q-2$.

Turning to the induction step, given an integer $\ell > 2$,
let $\code$ be a code of length $2\ell$ and size $M(\ell)$
that satisfies the conditions of the lemma.
Let the set $\code_1$ consist of all codewords $\bldc$ in $\code$ 
with the property that no codeword in $\code \setminus \{ \bldc \}$
agrees with $\bldc$ on its first $\ell{-}1$ positions.
Denote by $\code_2$ the complement set $\code \setminus \code_1$.

Let $\Subset$ be the set of all distinct
$(\ell{-}1)$-prefixes of the words in $\code_2$. 
No element in $\Subset$ can appear as an $(\ell{-}1)$-prefix
in any codeword in $\code_1$ and, so,
\[
|\code_1| \le q^{\ell-1} - |\Subset| \; .
\]
Since $\code$ has a decoder that detects
any single $\ell$-burst error, no two distinct words in $\code_2$
can have the same $\ell$-prefix, which means that
at most $q$ words in $\code_2$ can share the same $(\ell{-}1)$-prefix.
Hence,
\[
|\code_2| \le q \cdot |\Subset| 
\]
and, so,
\begin{eqnarray}
M(\ell) & = & |\code_1| + |\code_2| \nonumber \\
& \le &
|\code_1| + q \cdot |\Subset| \nonumber \\
& \le &
\label{eq:Mell}
(q^{\ell-1} - |\Subset|) + q \cdot |\Subset| \; .
\end{eqnarray}

For any element $v$ in the alphabet $F$ of $\code$, let $\code_2(v)$
denote the set of all codewords in $\code_2$ that end with $v$.
There exists at least one element $v' \in F$ for which
\[
|\code_2(v')| \ge \frac{|\code_2|}{q} = \frac{M(\ell) - |\code_1|}{q} 
\; .
\]
Let the mapping $\varphi : \code_2(v') \rightarrow F^{2\ell-2}$ be 
defined by
\[
\varphi(x_1 \, x_2 \, \ldots \, x_{2\ell-1} \, v') =
x_1 \, x_2 \, \ldots \, x_{\ell-1} \,
x_{\ell+1} \, x_{\ell+2} \, \ldots \, x_{2\ell-1} \, ;
\]
namely, $\varphi(\cdot)$ deletes (punctures) the entries of
its argument at the $\ell$th and $(2\ell)$th positions.
Denote by $\code'$ the set of images of this mapping:
\[
\code' =
\left\{ \varphi(\bldc) \;:\; \bldc \in \code_2(v') \right\} \; .
\]
Since $\code_2(v')$ satisfies
condition~\ref{item:bound2} for $\tau = \ell$,
then $\code'$ has to satisfy that condition for $\tau = \ell{-}1$;
furthermore, $\varphi(\cdot)$ is bijective and, so,
\[
|\code'| = |\code_2(v')| \ge
\frac{M(\ell) - |\code_1|}{q} \; ,
\]
or
\begin{eqnarray*}
M(\ell) & \le & |\code_1| + q \cdot |\code'| \\
& \le &
(q^{\ell-1} - |\Subset|) + q \cdot |\code'| \; .
\end{eqnarray*}
Combining the latter inequality with~(\ref{eq:Mell}) we thus get
\begin{equation}
\label{eq:code'1}
M(\ell) \le
(q^{\ell-1} - |\Subset|) +
q \cdot \min \left\{ |\Subset|, |\code'| \right\} \; .
\end{equation}

Next, we show that $\code'$ has
an $(\ell{-}1,\ell{-}1)$-burst list decoder.
If this were not the case, then there would be a word
\[
\bldy' =
y_1 \, y_2 \, \ldots \, y_{\ell-1} \,
y_{\ell+1} \, y_{\ell+2} \, \ldots \, y_{2\ell-1}
\]
in $F^{2\ell-2}$ and respective $\ell$ words
$\bldc'_0, \bldc'_1, \ldots, \bldc'_{\ell-1}$ in $\code'$ that would
form the violating configuration shown in
Figure~\ref{fig:badcodewords'}.

The respective pre-images $\bldc_i = \varphi^{-1}(\bldc'_i)$,
all belonging to $\code_2$ (and hence to $\code$),
would then look like the first $\ell$ rows in
the configuration of Figure~\ref{fig:badcodewords}
(with each block $\bldy_i$ therein replaced by
the element $y_i$ of $F$).
Recall, however, that since $\code_2$ is
the complement set of $\code_1$, each codeword in $\code_2$
agrees on the first $\ell{-}1$ positions with at least
one other codeword in $\code_2$. In particular,
there is a codeword $\bldc_\ell \in \code_2$
that agrees with
$\bldc_{\ell-1} \; (= \varphi^{-1}(\bldc'_{\ell-1}))$
on its first $\ell{-}1$ positions.
The codeword $\bldc_\ell$ could therefore serve as the last row
in Figure~\ref{fig:badcodewords}, thereby contradicting the fact
that $\code$ has an $(\ell,\ell)$-burst list decoder.
We conclude that $\code'$ has
an $(\ell{-}1,\ell{-}1)$-burst list decoder and, so,
\[
|\code'| \le M(\ell{-}1) \; .
\]

Combining the latter inequality with~(\ref{eq:code'1}) we get
\begin{eqnarray*}
M(\ell) & \le &
(q^{\ell-1} - |\Subset|) +
q \cdot \min \left\{ |\Subset|, M(\ell{-}1) \right\} \\
& \le &
\max_{t \in \Zring}
\left\{
(q^{\ell-1} - t) +
q \cdot \min \left\{ t, M(\ell{-}1) \right\} \right\} \\
& = &
q^{\ell-1}  + (q{-}1) \cdot M(\ell{-}1) \; .
\end{eqnarray*}
The result now follows by the induction hypothesis on
$M(\ell{-}1)$.\qed

\begin{theorem}
\label{thm:Reigerboundeveryell}
Let $\code$ be a code of length $n$ over an alphabet of size $q \ge 2$
and let $\ell$ and $\tau$ be positive integers that satisfy
the following three conditions:
\begin{enumerate}
\item
$\ell \,|\, \tau$, $\ell > 1$,  and $2\tau \le n$.
\item
There is a decoder for $\code$ that detects
any single $\tau$-burst error.
\item
There is an $(\ell,\tau)$-burst list decoder for $\code$.
\end{enumerate}
Then the redundancy $r$ of $\code$ satisfies
\[
r > \Bigl( 1 + \frac{1}{\ell} \Bigr) \tau - \log_q \ell \; .
\]%
\Endspace
\end{theorem}

\Proof{}
Denote by $F$ the alphabet of $\code$,
and let $\code'$ be defined as in the proof of
Theorem~\ref{thm:Reigerbound'}. Then $\code'$ is a code of length
$2 \tau$ over $F$ which satisfies
conditions~\ref{item:bound2}--\ref{item:bound3} and
\begin{equation}
\label{eq:code'tau=ell}
|\code'| \ge \frac{|\code|}{q^{n-2\tau}} \; .
\end{equation}
Write $b = \tau/\ell$.
By grouping together non-overlapping $b$-blocks over $F$,
we now regard $\code'$ as a code of length $2\ell$ over $F^b$.
As such, $\code'$ satisfies the conditions of
Lemma~\ref{lem:Reigerboundtau=ell} for an alphabet of size $q^b$.
Hence,
\[
|\code'| < \ell \cdot q^{b(\ell-1)} \; ,
\]
which readily implies with~(\ref{eq:code'tau=ell}) that
\[
|\code| < \ell \cdot q^{n-2\tau+b(\ell-1)} =
\ell \cdot q^{n-b(\ell+1)} 
\; .
\]
Thus, the redundancy $r$ of $\code$ satisfies
\begin{eqnarray*}
r
& = & n - \log_q |\code| \\
& > & n - \log_q (\ell \cdot q^{n-b(\ell+1)}) \\
& = &
b(\ell{+}1) - \log_q \ell \\
& = &
\Bigl( 1 + \frac{1}{\ell} \Bigr) \tau - \log_q \ell \; ,
\end{eqnarray*}
as claimed.\qed

In all our bounds, we have assumed that the code $\code$ has
a decoder that detects any single $\tau$-burst error
(condition~\ref{item:bound2} in all theorems).
We have also mentioned in Remark~\ref{rem:linear} that
when $\code$ is linear and $\ell < q$,
then condition~\ref{item:bound2}
is actually implied by condition~\ref{item:bound3}.
One could therefore ask whether condition~\ref{item:bound2}
is at all necessary in order to obtain our bounds.
The next example answers this question affirmatively:
it exhibits a code that does not satisfy
condition~\ref{item:boundell=2} and it violates
the bound of Proposition~\ref{prop:Reigerboundell=2}.

\begin{example}
\label{ex:n=4,tau=ell=2'}
Let $F$ be an alphabet of size $q \ge 2$,
select $\delta$ to be a nonzero element in $F$,
and consider the code $\code$ of length $4$ and size $2q$ over $F$
which is defined as the union of the following two sets:
\[
\Bigl\{ (a \, 0 \, 0 \, a) \;:\; a  \in F \Bigr\}
\; \phantom{.}
\]
and
\[
\Bigl\{ (a \, \delta \, \delta \, a) \;:\; a  \in F \Bigr\}
\; .
\]
We show that $\code$ has a $(2,2)$-burst list decoder
(while obviously, there is no decoder for $\code$
that can detect any single $2$-burst error).
Suppose to the contrary that there exist three distinct codewords
$\bldc_0, \bldc_1, \bldc_2 \in \code$ and respective three $2$-bursts
$\blde_0, \blde_1, \blde_2 \in F^4$ such that
\[
\bldc_0 + \blde_0 = \bldc_1 + \blde_1 = \bldc_2 + \blde_2 \; .
\]
Since no two codewords in $\code$ share the same $2$-suffix,
there can be at most one $2$-burst---say $\blde_0$---whose last
two entries are zero. By symmetry, $\blde_2$ (say) is the only
$2$-burst whose first two entries are zero.
Thus, $\blde_1$ can be zero only in its first and last positions,
which brings us to the configuration of Figure~\ref{fig:badcodewords};
namely, $\bldc_0$ and $\bldc_2$ are distinct right and left
neighbors of $\bldc_1$
(see the proof of Proposition~\ref{prop:Reigerboundell=2}).
However, this is impossible, since each codeword in $\code$
has exactly one neighbor (which is both a left neighbor and
a right neighbor).

Note that the redundancy of $\code$ equals
$4 - \log_q (2q) = 3 - \log_q 2$,
which is smaller than
the lower bound that we get for $\tau = 2$
in Proposition~\ref{prop:Reigerboundell=2}.\qed
\end{example}

The code in Example~\ref{ex:n=4,tau=ell=2'}
attains the next bound.

\begin{proposition}
\label{prop:withoutcondition2}
Let $\code$, $q$, and $\tau$ be as
in Proposition~\ref{prop:Reigerboundell=2},
except that $\code$ is not required
to satisfy condition~\ref{item:boundell=2}.
Then the redundancy $r$ of $\code$ satisfies
\[
        \left(
          1 + \frac{1}{2}
        \right) \, \tau - \log_q 2
       \; .
\]%
\Endspace
\end{proposition}

\Proof{}
We follow the steps of the proof of
Proposition~\ref{prop:Reigerboundell=2},
except that~(\ref{eq:contradictingell=2}) is replaced by
\[
|\code| = q^{n-r} > 2 q^{n-2\tau+b} \; .
\]
and~(\ref{eq:code'}) by
\[
|\code'| \ge 2q^b + 1 \; .
\]
Let $\code'_0$ be the set of all codewords in $\code'$
that have a right neighbor which is also a left neighbor.
By condition~\ref{item:bound3}, each codeword in $\code'_0$ has exactly
one such neighbor (which, obviously, is also an element of $\code'_0$).
Also, no codeword in $\code'_0$ can have
an ordinary neighbor (left or right) in $\code' \setminus \code'_0$,
(or else we would get
the violating configuration of Figure~\ref{fig:badcodewords}).
In particular, no $b$-suffix (respectively, $b$-prefix) of
a codeword in $\code'_0$ can appear as such in a codeword
that belongs to either $\code'_1$ or $\code'_2$
(where $\code'_1$ and $\code'_2$ are as in
the proof of Proposition~\ref{prop:Reigerboundell=2}).
Therefore,
\[
|\code'_1|, |\code'_2| \le q^b - \frac{|\code'_0|}{2}
\]
and, so,
\[
\left| \code' \setminus (\code'_0 \cup \code'_1 \cup \code'_2) \right|
\ge (2q^b + 1) - |\code'_0| -
2 \Bigl( q^b - \frac{|\code'_0|}{2} \Bigr) \ge 1 \; .
\]
We conclude that $\code'$ contains a codeword $\bldc_1$ that has
a right neighbor $\bldc_0$ and a left neighbor $\bldc_2$,
and these neighbors are distinct. But this brings us again
to the configuration in Figure~\ref{fig:badcodewords},
thereby reaching a contradiction.\qed

The example presented in Appendix~\ref{sec:appA} shows
that, in general, Theorem~\ref{thm:Reigerboundeveryell}
and Proposition~\ref{prop:withoutcondition2}
no longer hold if we omit from condition~\ref{item:bound1}
the assumption that $\tau$ is an integer multiple of $\ell$.

\section{Generalized Resultant of Certain Polynomials}
\label{sec:tools}

This section develops the tools that will be used in
Section~\ref{sec:RSbound}
to show that Reed--Solomon codes attain
the bound~(\ref{eq:Reigerboundlinear}).
In particular, Theorem~\ref{thm:determinant} below presents
a generalization of the known formula for the resultant
of two polynomials, to a larger number of polynomials that have
a certain structure.

For a field $F$ and an integer $k$, denote by $F_k[x]$
the set of all polynomials over $F$ of degree less than $k$
in the indeterminate $x$.

Let $F$ be the finite field $\GF(q)$ and let $r$ be a positive integer.
Fix $\alpha$ to be a nonzero element in $F$ with multiplicative order
at least $r$, and let $\bldbeta = (\beta_i)_{i=0}^\ell$
be a vector whose $\ell{+}1$ entries are all nonzero elements of $F$.
Let $\mu_0, \mu_1, \ldots, \mu_\ell$ be positive integers such that
\begin{equation}
\label{eq:mui}
\sum_{i=0}^\ell \mu_i = r \; .
\end{equation}
For $i = 0, 1, \ldots, \ell$, define
\[
\tau_i = r - \mu_i \; , \quad 0 \le i \le \ell \; ,
\]
and for an indeterminate $x$,
denote by $M_i(x;\beta_i)$ the expression
\[
M_i(x;\beta_i) = \prod_{j=0}^{\tau_i-1} (x - \beta_i \alpha^j) \; .
\]
We regard $M_i(x;\beta_i)$ as a univariate polynomial over $F$
in the indeterminate $x$, with $\beta_i$ serving as a parameter.

In this section, we prove the following result.

\begin{theorem}
\label{thm:tool}
The following two conditions are equivalent:
\begin{list}{}{\settowidth{\labelwidth}{\textit{(ii)}}}
\item[(i)]
There exist polynomials
\begin{equation}
\label{eq:poldeg}
u_i(x) \in F_{\mu_i}[x] , \quad 0 \le i \le \ell \; ,
\end{equation}
not all zero, such that
\begin{equation}
\label{eq:polsum}
\sum_{i=0}^\ell u_i(x) M_i(x;\beta_i) = 0 \; .
\end{equation}
\item[(ii)]
For some distinct $i$ and $k$ in the range
$0 \le \ i, k \ \le \ell$ and some integer $t$ in the range
$-\mu_i < t < \mu_k$,
\[
\frac{\beta_k}{\beta_i} = \alpha^t \; .
\]%
\Endspace
\end{list}
\end{theorem}

\Proof{}
This theorem is implied by the considerations in the following
paragraphs, in particular by Theorem~\ref{thm:determinant}.\qed

For each $i = 0, 1, 2, \ldots, \ell$, write
\[
M_i(x;\beta_i) = \sum_{j=0}^{\tau_i} M_{i,j} x^j
\]
(where $M_{i,j}$ is a function of $\beta_i$), and define $A_i(\beta_i)$
to be the following $\mu_i \times r$ echelon matrix
over $F$:
\begin{equation}
\label{eq:Ai}
\newcommand{\bigzero}{\textrm{\huge{0}}}
A_i(\beta_i) =
\left(
\arraycolsep0.75ex
\begin{array}{cccccccc}
M_{i,0} & M_{i,1} & \ldots & M_{i,\tau_i} &              &        & \\
        & M_{i,0} & M_{i,1}& \ldots       & M_{i,\tau_i} &        &
\multicolumn{1}{l}{\bigzero} \\
\multicolumn{1}{r}{\bigzero}
        &         & \ddots & \ddots       & \cdots       & \ddots & \\
        &         &        & M_{i,0} & M_{i,1} & \ldots & M_{i,\tau_i}
\end{array}
\right)
\; .
\end{equation}
Then, (\ref{eq:poldeg})--(\ref{eq:polsum}) can be expressed
in matrix form as
\[
\sum_{i=0}^\ell \bldu_i A_i(\beta_i) = \bldzero \; ,
\]
where each $\bldu_i$ is a row vector in $F^{\tau_i}$,
and at least one of these vectors is nonzero.
Equivalently,
\[
\bldu A = \bldzero \; ,
\]
where $\bldu$ is a nonzero vector in $F^r$
and $A = A(\bldbeta)$ is the following $r \times r$ matrix over $F$:
\[
A(\bldbeta) =
\left(
\renewcommand{\arraystretch}{1.5}
\begin{array}{c}
A_0(\beta_0) \\ \hline A_1(\beta_1) \\ \hline \vdots \\
\hline A_\ell(\beta_\ell)
\end{array}
\right)
\; .
\]

\begin{theorem}
\label{thm:determinant}
\textup{(Generalized resultant of $M_i(x;\beta_i)$)}
For some nonzero constant $\kappa(\alpha) \in F$
(which depends on $\alpha$ but not on $\bldbeta$),
\begin{equation}
\label{eq:determinant}
\det(A(\bldbeta)) =
\kappa(\alpha) \cdot
\prod_{0 \le i < k \le \ell}
\prod_{s=0}^{\mu_i-1}
\prod_{t=0}^{\mu_k-1}
(\beta_k \alpha^s - \beta_i \alpha^t)
\; .
\end{equation}%
\Endspace
\end{theorem}

To prove the latter theorem,
we regard $\bldbeta$ as a vector of indeterminates and
\[
\Delta(\bldbeta) = \det(A(\bldbeta))
\]
as a multivariate polynomial over $F$.
The properties of this polynomial are summarized
in Lemmas~\ref{lem:nonzero}--\ref{lem:factor} below,
and Theorem~\ref{thm:determinant} will then follow as
a direct corollary of these properties.

Given a vector $\bldxi = (\xi_0 \, \xi_1 \, \ldots \, \xi_{m-1})$,
we denote by $V(\bldxi)$ the $m \times m$ Vandermonde matrix
\[
V(\bldxi) = \left( \, \xi_t^s \, \right)_{s,t=0}^{m-1}
            \; .
\]
We will use the notation $V_m$ for
$V(1 \, \alpha \, \alpha^2 \, \ldots \, \alpha^{m-1})$.

\begin{lemma}
\label{lem:nonzero}
The multivariate polynomial $\Delta(\bldbeta)$ is not identically zero.
\end{lemma}

\Proof{}
We find an assignment
$\bldbeta^* = (\beta_i^*)_{i=0}^\ell$
for $\bldbeta$ for which $\Delta(\bldbeta^*) \ne 0$.
For $i = 0, 1, \ldots, \ell$, define the partial sums
\[
r_i = \mu_0 + \mu_1 + \cdots + \mu_i
\]
and
\[
\beta_i^* = \alpha^{r_i} \; .
\]
Taking the product of $A_i(\beta_i^*)|_{\beta_i^* = \alpha^{r_i}}$ and
$V_r$, one can check that the nonzero columns of
the resulting $\mu_i \times r$ matrix
$A_i(\alpha^{r_i}) V_r$ are indexed by integers $j$ in the range
$0 \le j < r_i$. Furthermore, the $\mu_i$ columns that are indexed by
\[
r_{i-1} \le j <  r_i
\]
(with $r_{-1} = 0$) form a $\mu_i \times \mu_i$ nonsingular matrix $X_i$
which is obtained by multiplying a Vandermonde matrix to the right
by a diagonal matrix; specifically:
\begin{equation}
\label{eq:Xi}
X_i =
\Bigl(
\begin{array}{c}
\alpha^{(r_{i-1} + t)s}
\end{array}
\Bigr)_{s,t=0}^{\mu_i-1}
\cdot
\textrm{diag}
\left( M_i(\alpha^{r_{i-1} + t};\alpha^{r_i}) \right)_{t=0}^{\mu_i-1}
\; .
\end{equation}
It follows that the respective matrix
$A(\bldbeta^*) V_r$ has a block-triangular form and, so,
\begin{eqnarray}
\Delta(\bldbeta^*) =
\det(A(\bldbeta^*)) & = &
\frac{\det(A(\bldbeta^*) V_r)}{\det(V_r)} \nonumber \\
& = &
\frac{1}{\det(V_r)}\prod_{i=0}^\ell \det(X_i) \nonumber \\
\label{eq:dethatA}
& \ne & 0 \; .
\end{eqnarray}
\qed

\begin{lemma}
\label{lem:degree}
For each $i = 0, 1, \ldots, \ell$,
the degree of $\beta_i$ in $\Delta(\bldbeta)$ is at most $\mu_i \tau_i$.
\end{lemma}

\Proof{}
By inspecting the matrix $A(\bldbeta)$ we see that
the largest contribution to the degree of $\beta_i$ can be made by
the leftmost (main) diagonal in $A_i(\beta_i)$:
the product of the elements along that diagonal is
\[
M_{i,0}^{\mu_i} = (-\alpha^{(\tau_i-1)/2} \beta_i)^{\mu_i \tau_i}
\; ,
\]
and, so, the degree of $\beta_i$ in $\Delta(\bldbeta)$ can be
at most $\mu_i \tau_i$.\qed

\begin{lemma}
\label{lem:factor}
For every distinct $i, k \in \{ 0, 1, \ldots, \ell \}$,
the multivariate polynomial $\Delta(\bldbeta)$ is divisible by
\[
\prod_{t=0}^{\mu_k-1}
(\beta_k - \beta_i \alpha^t)^{\min \{ \mu_i, \mu_k - t \}} \; .
\]%
\Endspace
\end{lemma}

\Proof{}
Due to symmetry, it suffices to prove the lemma assuming $i = 0$.
Hereafter in this proof, we fix $k$ to be some element in
$\{ 1, 2, \ldots, \ell \}$.
While it is not too difficult to see that $\beta_k - \beta_0 \alpha^t$
is a factor of $\Delta(\bldbeta)$, we also need to establish the
multiplicity of that factor.
We do this by introducing $\mu_0$ new
indeterminates which are given by the entries of the following vector
$\bldgamma$:
\[
\bldgamma = ( \gamma_h )_{h=0}^{\mu_0-1} \; .
\]
We define the respective polynomials
\[
\sigma_h(x;\gamma_h) =
\prod_{j=0}^{\tau_0+h-1} (x - \gamma_h \alpha^j) \; ,
\quad
0 \le h < \mu_0 \; ,
\]
and regard them as univariate polynomials in
the indeterminate $x$ over the field
\[
\Phi =
F(\beta_1{,}\beta_2{,}\ldots{,}\beta_{k-1}{,}
\beta_{k+1}{,}\beta_{k+2}{,}\ldots{,}\beta_\ell{,}
\gamma_0{,}\gamma_1{,}\ldots{,}\gamma_{\mu_0-1}) ;
\]
namely, $\Phi$ is the rational function field over $F$ where the
indeterminates are all the entries of $\bldbeta$ and $\bldgamma$,
except for $\beta_k$.
(The analysis in the sequel will involve
univariate polynomials in the indeterminate $\beta_k$ over $\Phi$,
as well as the rational function field $\Phi(\beta_k)$.)
Notice that when we substitute $\gamma_h = \beta_0$, we get
\begin{equation}
\label{eq:sigma0M0}
\sigma_h(x;\beta_0) =
M_0(x;\beta_0) \cdot
\prod_{j=0}^{h-1}(x - \beta_0 \alpha^{\tau_0+j}) \; .
\end{equation}

Let $S_0(\bldgamma)$ be the $\mu_0 \times r$ matrix
over $\Phi$ whose rows are given by the coefficients of
$\sigma_h(x;\gamma_h)$, for $0 \le h < \mu_0$
(i.e., entry $(h,j)$ in $S_0(\bldgamma)$ is the coefficient
of $x^j$ in $\sigma_h(x;\gamma_h)$).
It follows from~(\ref{eq:sigma0M0}) that
when we substitute
$\bldgamma = \bldbeta_0 = (\beta_0 \, \beta_0 \, \ldots \, \beta_0)$,
then
$S_0(\bldbeta_0)$ and $A_0(\beta_0)$ are related by
\begin{equation}
\label{eq:S0A0}
S_0(\bldbeta_0) = L A_0(\beta_0) \; ,
\end{equation}
where $L$ is a $\mu_0 \times \mu_0$ lower-triangular matrix
having $1$'s along its main diagonal.

Let $S(\beta_k;\bldgamma)$ be the following $r \times r$ matrix
over the field $\Phi(\beta_k)$:
\[
S(\beta_k;\bldgamma) =
S(\beta_1,\beta_2,\ldots,\beta_\ell;\bldgamma) =
\left(
\renewcommand{\arraystretch}{1.5}
\begin{array}{c}
S_0(\bldgamma) \\
\hline A_1(\beta_1) \\
\hline A_2(\beta_2) \\ \hline \vdots \\
\hline A_\ell(\beta_\ell)
\end{array}
\right)
\; .
\]
  From~(\ref{eq:S0A0}) we get that, in $\Phi(\beta_k)$,
\begin{equation}
\label{eq:detS}
\Delta(\bldbeta) =
\det(S(\beta_k;\bldbeta_0)) \; .
\end{equation}

Let $f(\beta_k;\bldgamma)$ be the following univariate polynomial
in the indeterminate $\beta_k$ over $\Phi$:
\begin{equation}
\label{eq:f}
f(\beta_k;\bldgamma) =
\det(S(\beta_k;\bldgamma)) \; .
\end{equation}
We verify that for every $0 \le t < \mu_k$ and every $h$ in the range
\[
\Range(t) =
\Bigl\{
h \;:\;
\max \{ 0, t + \mu_0 - \mu_k \} \le h < \mu_0
\Bigr\} \; ,
\]
the element $\gamma_h \alpha^t$ is a root of $f(\beta_k;\bldgamma)$.  We do
this by demonstrating that for any such $t$ and $h$, the rows of $S(\gamma_h
\alpha^t;\bldgamma)$ are linearly dependent over $\Phi$.  Specifically, we
exhibit nonzero $\blde \in \Phi^{\mu_0}$ and $\bldu \in \Phi^{\mu_k}$ such
that
\begin{equation}
\label{eq:uv}
\blde S_0(\bldgamma) - \bldu A_k(\gamma_h \alpha^t) =
\bldzero \; .
\end{equation}
Given $t$ and $h$, let $u(x)$ be the following univariate polynomial
over $\Phi$:
\[
u(x) =
\left( \prod_{j=0}^{t-1} (x - \gamma_h \alpha^j) \right)
\left( \prod_{j=t}^{h+\mu_k-\mu_0-1}
(x - \gamma_h \alpha^{\tau_k+j}) \right) \; .
\]
Since $h \in \Range(t)$ we have
\[
\deg u(x) = h + \mu_k - \mu_0 < \mu_k \; ,
\]
so we can take $\bldu$ to be the vector of coefficients of $u(x)$.
We readily get that
\begin{eqnarray*}
u(x) M_k(x;\gamma_h \alpha^t) & =  &
\prod_{j=0}^{\tau_k+h+\mu_k-\mu_0-1} (x - \gamma_h \alpha^j) \\
& = &
\prod_{j=0}^{\tau_0+h-1} (x - \gamma_h \alpha^j) =
\sigma_h(x;\gamma_h) \; .
\end{eqnarray*}
Hence, (\ref{eq:uv}) holds when $\blde$ is taken as
a unit vector having $1$ at position $h$.

We conclude that, over $\Phi$, the polynomial $f(\beta_k;\bldgamma)$
is divisible by
\[
\prod_{t=0}^{\mu_k-1} \prod_{h \in \Range(t)}
(\beta_k - \gamma_h \alpha^t)
\; .
\]
Substituting $\bldgamma = \bldbeta_0$, it follows that
$f(\beta_k;\bldbeta_0)$ is divisible by
\[
\prod_{t=0}^{\mu_k-1}
(\beta_k - \beta_0 \alpha^t)^{|\Range(t)|} =
\prod_{t=0}^{\mu_k-1}
(\beta_k - \beta_0 \alpha^t)^{\min \{ \mu_0, \mu_k - t \}} \; ,
\]
and, by~(\ref{eq:detS})--(\ref{eq:f}), so is $\Delta(\bldbeta)$.
\qed

\Proof{of Theorem~\ref{thm:determinant}}
The right-hand side of~(\ref{eq:determinant})
factors over $F$ as follows:
for every distinct $i, k \in \{ 0, 1, \ldots, \ell \}$
and every $0 \le t < \mu_k$, the term
\[
\beta_k - \beta_i \alpha^t
\]
has multiplicity $\min \{ \mu_i, \mu_k - t \}$
in the right-hand side of~(\ref{eq:determinant})
(for $t = 0$, we regard $\beta_k - \beta_i$ and
$\beta_i - \beta_k$ as the same term).
By Lemma~\ref{lem:factor} we then get that
the right-hand side of~(\ref{eq:determinant})
divides $\Delta(\bldbeta)$.
Furthermore, for each $\{ 0, 1, \ldots, \ell \}$,
the degree of $\beta_i$ in
the right-hand side of~(\ref{eq:determinant}) equals
\[
\sum_{k=0 \atop k \ne i}^\ell \mu_i \mu_k =
\mu_i (r - \mu_i) = \mu_i \tau_i \; .
\]
Hence, by Lemmas~\ref{lem:nonzero} and~\ref{lem:degree}
we conclude that
that the right-hand side of~(\ref{eq:determinant})
actually equals $\Delta(\bldbeta)$.

The exact expression for $\kappa(\alpha)$ is given
in Appendix~\ref{sec:appB}.\qed

\begin{remark}
For $\ell = 1$ (in which case $\tau_0 + \tau_1 = r$),
the matrix $A(\bldbeta)$ is
the Sylvester matrix~\cite{Cox:Little:OShea:97:1} of the polynomials
$M_0(x;\beta_0)$ and $M_1(x;\beta_1)$
(up to reversal of the order of the rows and columns),
and Theorem~\ref{thm:determinant} then provides
the known formula for the resultant of
these polynomials~\cite[p.~36]{Lidl:Niederreiter:97:1}.\qed
\end{remark}

\begin{remark}
For $r = \ell{+}1$ (in which case $\mu_i = 1$ for all $i$),
the matrix $A(\bldbeta)$ is related to
the $r \times r$ Vandermonde matrix $V(\bldbeta)$ by
\[
A(\bldbeta) = V^{\mathsf{T}}(\bldbeta) U(\alpha)
\; ,
\]
where $V^{\mathsf{T}}(\bldbeta)$ is the transpose of $V(\bldbeta)$ and
where $U(\alpha)$ does not depend on $\bldbeta$ and is zero below
its main anti-diagonal.
Theorem~\ref{thm:determinant} then provides the known formula
for the determinant of a square Vandermonde matrix.\qed
\end{remark}

\section{Burst List Decoding of Reed--Solomon Codes}
\label{sec:RSbound}

The goal of this section is to show that the well-known
Reed--Solomon codes achieve the generalized Reiger bound for
linear codes
(see~Equation~(\ref{eq:Reigerboundlinear}) in Remark~\ref{rem:linear}).

Let $F$ be the finite field $\GF(q)$ and let $\alpha$ be
an element of multiplicative order $n$ in $F$.
For a nonnegative integer $r < n$, denote by
$\code_\RS(n,r)$ the $[n,k{=}n{-}r]$ Reed--Solomon code over $F$
with a parity-check matrix
\[
H_\RS = 
\left( \, \alpha^{sj} \, \right)_{s=0,\;j=0}^{r-1 \;\; n-1}
\; .
\]

The following theorem shows that $\code_\RS(n,r)$ attains
the bound~(\ref{eq:Reigerboundlinear}).

\begin{theorem}
\label{thm:RSbound}
Let $\ell$ and $\tau$ be positive integers such that
\begin{equation}
\label{eq:RSbound}
r \ge \tau + \left\lceil \frac{\tau}{\ell} \right\rceil \; .
\end{equation}
Then there is an $(\ell,\tau)$-burst list decoder for $\code_\RS(n,r)$.
\end{theorem}

\Proof{} We will assume in the proof that~(\ref{eq:RSbound}) holds with
equality; otherwise, just reduce $r$ to the right-hand side
of~(\ref{eq:RSbound}). Recalling the coset characterization of $\tau$-burst
errors in Section~\ref{sec:gen:reiger:bound:group:codes:1}, we suppose to the
contrary that there exist $\ell{+}1$ distinct row vectors $\blde_0, \blde_1,
\ldots, \blde_\ell \in F^n$ such that
\begin{equation}
\label{eq:samesyndrome}
H_\RS \blde_0^T = H_\RS \blde_1^T = \cdots = H_\RS \blde_\ell^T \; ,
\end{equation}
where the support of each $\blde_i$ is contained in a subset
\[
J_i = \{ \lambda_i + t \;:\; 0 \le t < \tau \} \; ;
\]
here each $\lambda_i$ is an integer in the range
$0 \le \lambda_i \le n-\tau$.
We observe that since the minimum distance of $\code_\RS(n,r)$
is $r{+}1$,
for every distinct $i, k \in \{ 0, 1, \ldots, \ell \}$
we must have
\[
|J_i \cup J_k| > r \; ,
\]
which readily implies that for $i \ne k$,
\[
|J_i \setminus J_k| > r - \tau =
\left\lceil \frac{\tau}{\ell} \right\rceil \; .
\]
Thus, for every distinct $i, k \in \{ 0, 1, \ldots, \ell \}$,
\begin{equation}
\label{eq:lambda}
\| \lambda_k - \lambda_i \|_n >
\left\lceil \frac{\tau}{\ell} \right\rceil \; ,
\end{equation}
where
\[
\| a \|_n =
\left\{
\begin{array}{lcl}
|a|     && \textrm{if $0 \le |a| \le n/2$} \\
n - |a| && \textrm{if $n/2 < |a| < n$}
\end{array}
\right.
\; .
\]

The sum of the sizes of the sets $J_i$ is $(\ell{+}1) \tau$,
and this value may be smaller than $\ell r$ in case $\tau$
is not divisible by $\ell$. For convenience in the sequel,
we will now artificially expand some of the sets $J_i$ by one,
by adding the element $\lambda_i + \tau$,
so that the sum of the sizes becomes exactly $\ell r$.
Letting $\tau_i$ be the size of
(the possibly expanded) $J_i$ and defining
\[
\mu_i = r - \tau_i \; ,
\]
we have
\[
\sum_{i=0}^\ell \mu_i = 
\sum_{i=0}^\ell (r - \tau_i) =
(\ell{+}1) r - \sum_{i=0}^\ell \tau_i = r
\]
(see~(\ref{eq:mui})).

Denote by $H_i$
the $r \times \tau_i$ sub-matrix of $H_\RS$ which is formed
by the columns of $H$ that are indexed by $J_i$, namely:
\[
H_i =
\left( \, \alpha^{(\lambda_i+t)s} \,
\right)_{s=0,\;t=0}^{r-1 \;\; \tau_i-1}
\; .
\]
Define the $r \times r$ matrix $T_i$ by
\[
T_i =
\left(
\renewcommand{\arraystretch}{1.2}
\begin{array}{c|c}
\makebox[2ex]{$I_i$} & \makebox[2ex]{$0$} \\
\hline
\multicolumn{2}{c}{A_i(\alpha^{\lambda_i})}
\end{array}
\right)
\; ,
\]
where $I_i$ is a $\tau_i \times \tau_i$ identity matrix
and $A_i(\cdot)$ is given by~(\ref{eq:Ai}).
Notice that $A_i(\alpha^{\lambda_i}) H_i = 0$
and, so, the product $T_i H_i$
results in an $r \times \tau_i$ matrix $Y_i$ which takes
the following form:
\begin{equation}
\label{eq:Yi}
Y_i = T_i H_i =
\left(
\renewcommand{\arraystretch}{1.5}
\begin{array}{c}
\left( \, \alpha^{(\lambda_i+t)s} \, \right)_{s,t=0}^{\tau_i-1}
\\
\hline
0
\end{array}
\right) \; .
\end{equation}
Specifically, the first $\tau_i$ rows of this matrix form
a nonsingular square Vandermonde matrix, whereas the remaining $\mu_i$
rows are all zero.

Consider the following $\ell r \times \ell r$ matrix $B$:
\[
\newcommand{\matblock}[1]{\multicolumn{1}{|c|}{\makebox[2ex]{$#1$}}}
B = 
\left(
\;
\renewcommand{\arraystretch}{2.0}
\newcommand{\bigzero}{\textrm{\huge{0}}}
\arraycolsep2.5ex
\begin{array}{ccccc}
\cline{1-2}
\matblock{H_0} & \matblock{-H_1} && \multicolumn{2}{c}{\bigzero}     \\
\cline{1-3}
\matblock{H_0} && \matblock{-H_2} &&                                 \\
\cline{1-1}\cline{3-3}
\vdots         &&& \ddots          &                                 \\
\cline{1-1}\cline{5-5}
\matblock{H_0} &  \multicolumn{2}{c}{\bigzero} && \matblock{-H_\ell} \\
\cline{1-1}\cline{5-5}
\end{array}
\;
\right)
\; .
\]
Next, we multiply $B$ to the left by an $\ell r \times \ell r$
block-diagonal matrix $T$ which contains
the blocks $T_1, T_2, \ldots, T_\ell$ along its main diagonal:
\[
\newcommand{\matblock}[1]{\multicolumn{1}{|c|}{\makebox[2ex]{$#1$}}}
T B = 
\left(
\;
\renewcommand{\arraystretch}{2.0}
\newcommand{\bigzero}{\textrm{\huge{0}}}
\arraycolsep2.5ex
\begin{array}{ccccc}
\cline{1-2}
\matblock{Z_1} & \matblock{-Y_1} && \multicolumn{2}{c}{\bigzero}     \\
\cline{1-3}
\matblock{Z_2} && \matblock{-Y_2} &&                                 \\
\cline{1-1}\cline{3-3}
\vdots         &&& \ddots          &                                 \\
\cline{1-1}\cline{5-5}
\matblock{Z_\ell}&\multicolumn{2}{c}{\bigzero} && \matblock{-Y_\ell} \\
\cline{1-1}\cline{5-5}
\end{array}
\;
\right)
\; ,
\]
where $Y_i$ is given by~(\ref{eq:Yi}) and
\[
Z_i = T_i H_0 =
\left(
\renewcommand{\arraystretch}{1.5}
\begin{array}{c}
\left( \, \alpha^{(\lambda_0+t)s} \, \right)_{s,t=0}^{\tau_0-1}
\\
\hline
A_i(\alpha^{\lambda_i}) H_0
\end{array}
\right) \; .
\]
Our contradicting assumption~(\ref{eq:samesyndrome}) implies that $B$
has dependent columns and is therefore singular.
This, in turn, implies the singularity of
the $\tau_0 \times \tau_0$ matrix
\[
\left(
\renewcommand{\arraystretch}{1.5}
\begin{array}{c}
A_1(\alpha^{\lambda_1}) H_0 \\ \hline
A_2(\alpha^{\lambda_2}) H_0 \\ \hline \vdots \\ \hline
A_\ell(\alpha^{\lambda_\ell}) H_0
\end{array}
\right)
\; ,
\]
which is formed by taking the last $\mu_i$ rows of each $Z_i$
and stacking them together for all $i = 1, 2, \ldots, \ell$
(notice that $\sum_{i=1}^\ell \mu_i = r - \mu_0 = \tau_0$).
Hence, there exist row vectors $\bldu_1, \bldu_2, \ldots, \bldu_\ell$,
not all zero, such that $\bldu_i \in F^{\mu_i}$ and
\[
\sum_{i=1}^\ell \bldu_i A_i (\alpha^{\lambda_i}) H_0  = \bldzero \; .
\]
Equivalently, there exist polynomials
\[
u_i(x) \in F_{\mu_i}[x] \; , \quad 1 \le i \le \ell \; ,
\]
not all zero, such that
\[
\sum_{i=1}^\ell
u_i(\alpha^{\lambda_0+t})
M_i(\alpha^{\lambda_0 + t}; \alpha^{\lambda_i}) = 0 \; ,
\quad
0 \le t < \tau_0 \; .
\]
But the latter condition means that the polynomial
\[
\sum_{i=1}^\ell u_i(x) M_i(x;\alpha^{\lambda_i})
\]
(which is in $F_r[x]$) is divisible by
$M_0(x;\alpha^{\lambda_0})$; namely,
there exists a $u_0(x) \in F_{\mu_0}[x]$ such that
\[
\sum_{i=0}^\ell u_i(x) M_i(x;\alpha^{\lambda_i}) = 0 \; .
\]
We then get from
Theorem~\ref{thm:tool} that there exist
distinct $i, k \in \{ 0, 1, \ldots, \ell \}$ such that
\[
\| \lambda_k - \lambda_i \|_n <
\max \{ \mu_i, \mu_k \} \le
\left\lceil \frac{\tau}{\ell} \right\rceil \; .
\]
This, however, contradicts~(\ref{eq:lambda}).\qed

\ifIEEE\appendices\else\appendix\section*{\centering{Appendices}}\fi

\section{Example}
\label{sec:appA}

We demonstrate here that, in general,
Theorems~\ref{thm:Reigerbound} and~\ref{thm:Reigerbound'}
no longer hold if condition~\ref{item:bound1} therein is relaxed
to requiring only that $2 \tau \le n$. Specifically, we
present an example of a linear code $\code$ of length $n = 8$
and redundancy $r = 4$ over $F = \GF(q)$
which satisfies conditions~\ref{item:bound2}
and~\ref{item:bound3} in the theorem for $\ell = 2$ and $\tau = 3$.
For these parameters, condition~\ref{item:bound1} in
Theorems~\ref{thm:Reigerbound} and~\ref{thm:Reigerbound'} is violated,
and the redundancy lower bounds in these theorems
indeed do not hold.
In particular,
the specialized redundancy lower
bound~(\ref{eq:Reigerboundlinear}) for linear codes in
Remark~\ref{rem:linear} does not hold either.

The code $\code$ is generated by the matrix
\[
G =
\left(
\arraycolsep0.25ex
\begin{array}{cccccccc}
1 & * & * & 0 & 1 & 0 & 0 & 0 \\
0 & 1 & * & 0 & 1 & 1 & 0 & 0 \\
0 & 0 & 1 & 1 & 0 & * & 1 & 0 \\
0 & 0 & 0 & 1 & 0 & * & * & 1
\end{array}
\right)
\; ,
\]
where the stars stand hereafter for arbitrary elements of $F$.
Since the rows of $G$ form a diagonal band of $5$-bursts,
it follows that none of the nonzero codewords of $\code$ is
a $4$-burst and, so, $\code$ satisfies condition~\ref{item:bound2}
of Theorem~\ref{thm:Reigerbound}. Furthermore, if
$\bldc_0 + \blde_0 = \bldc_1 + \blde_1$ for distinct codewords
$\bldc_0, \bldc_1 \in \code$ and nonzero $3$-burst errors $\blde_0$
and $\blde_1$, then the leftmost entries in $\blde_0$ and $\blde_1$
have to be at least two positions apart. (Similarly, the rightmost
entries in $\blde_0$ and $\blde_1$ have to be at least two positions
apart.) We next show that a violating configuration
\[
\bldc_0 + \blde_0 = \bldc_1 + \blde_1 = \bldc_2 + \blde_2
\]
cannot exist (for distinct $\bldc_0, \bldc_1, \bldc_2 \in \code$)
by distinguishing between several cases.

\emph{Case~1:}
Suppose to the contrary that there exists a violating configuration 
with error words of the form
\[
\arraycolsep0.25ex
\begin{array}{cccccccccc}
\blde_0 & = & ( * & * & * & 0 & 0 & 0 & 0 & 0 ) \\
\blde_1 & = & ( 0 & 0 & * & * & * & 0 & 0 & 0 ) \\
\blde_2 & = & ( 0 & 0 & 0 & 0 & * & * & * & 0 )
\end{array}
\; ,
\]
and assume without loss of generality that $\bldc_1 = \bldzero$.
Then, from $\bldc_0 - \bldc_1 = \blde_1 - \blde_0$
we deduce that $\bldc_0$ takes the form
\[
\arraycolsep0.25ex
\begin{array}{cccccccccc}
\bldc_0 & = & ( * & * & * & * & * & 0 & 0 & 0 )
\end{array}
\; ,
\]
which means that $\bldc_0$ has to be a nonzero scalar multiple 
of the first row of $G$.
Also, from $\bldc_2 - \bldc_1 = \blde_1 - \blde_2$ we get that
\[
\arraycolsep0.25ex
\begin{array}{cccccccccc}
\bldc_2 & = & ( 0 & 0 & * & * & * & * & * & 0 )
\end{array}
\; ,
\]
which means that $\bldc_2$ is a nonzero scalar multiple of
the third row in $G$. Therefore,
the fourth position in $\bldc_0$ is zero while it is nonzero in
$\bldc_2$, and this, in turn, implies that
the fourth position in $\bldc_0 - \bldc_2$ is nonzero also.
Yet, the latter contradicts the fact that
$\bldc_0 - \bldc_2 = \blde_2 - \blde_0$.

\emph{Case~2:}
Suppose now that the violating configuration takes the form
\[
\arraycolsep0.25ex
\begin{array}{cccccccccc}
\blde_0 & = & ( * & * & * & 0 & 0 & 0 & 0 & 0 ) \\
\blde_1 & = & ( 0 & 0 & * & * & * & 0 & 0 & 0 ) \\
\blde_2 & = & ( 0 & 0 & 0 & 0 & 0 & * & * & * )
\end{array}
\; \phantom{.}
\]
($\blde_0$ and $\blde_1$ are as in Case~1, yet
the support of $\blde_2$ is shifted one position to the right).
Assuming again that $\bldc_1 = \bldzero$,
we get that $\bldc_0$ has to be a nonzero scalar multiple of
the first row of $G$ while $\bldc_2$ has to be a nonzero
linear combination of the last two rows of $G$.
Hence, the fifth position in $\bldc_0 - \bldc_2$ cannot be zero,
yet this contradicts the fact that
$\bldc_0 - \bldc_2 = \blde_2 - \blde_0$.

There are two other violating configurations to consider,
which are obtained by reversing the order of coordinates in
the error patterns covered by Cases~1 and~2.
The proof of contradiction remains
the same due to the symmetries of $G$.

The code $\code$ also serves to demonstrate that for $q \ge 4$,
Theorem~\ref{thm:Reigerboundeveryell} becomes false if we remove from
condition~\ref{item:bound1} therein the assumption that $\tau$ is an integer
multiple of $\ell$. A similar statement holds for
Proposition~\ref{prop:withoutcondition2} and $q \ge 5$.

\section{Formula for $\kappa(\alpha)$}
\label{sec:appB}

For the sake of completeness,
we compute here the constant $\kappa(\alpha)$
which appears in the right-hand side of~(\ref{eq:determinant})
in Theorem~\ref{thm:determinant}.
We continue where we left off
in the proof of Lemma~\ref{lem:nonzero}
and obtain an expression for $\Delta(\bldbeta^*)$
using~(\ref{eq:dethatA}).

To this end, we first compute the determinant of
the matrix $X_i$ defined in~(\ref{eq:Xi}):
\begin{eqnarray*}
\det(X_i) & = &
\prod_{s=0}^{\mu_i-1} 
\Bigl(
M_i(\alpha^{r_{i-1} + s};\alpha^{r_i}) 
\Bigr. \\
&& \makebox[10ex]{} \cdot \Bigl.
\prod_{t=s+1}^{\mu_i-1} 
(\alpha^{r_{i-1} + t} - \alpha^{r_{i-1} + s})
\Bigr) \\
& = &
(-1)^{\mu_i \tau_i}
\prod_{s=0}^{\mu_i-1} 
\Bigl(
\prod_{t=0}^{\tau_i-1}
(\alpha^{r_i + t} - \alpha^{r_{i-1} + s})
\Bigr. \\
&& \makebox[10ex]{} \cdot
\prod_{t=s+1}^{\mu_i-1} 
(\alpha^{r_{i-1} + t} - \alpha^{r_{i-1} + s})
\Bigr) \\
& = &
(-1)^{\mu_i \tau_i}
\prod_{s=0}^{\mu_i-1} 
\prod_{t=s+1}^{r-1}
(\alpha^{r_{i-1} + t} - \alpha^{r_{i-1} + s}) \\
& = &
(-1)^{\mu_i (r - \mu_i)} \cdot
\alpha^{r_{i-1} \mu_i (r - (\mu_i+1)/2)} \\
&& \makebox[10ex]{} \cdot
\prod_{s=0}^{\mu_i-1} 
\prod_{t=s+1}^{r-1} (\alpha^t - \alpha^s) \; .
\end{eqnarray*}
Plugging the latter expression into~(\ref{eq:dethatA})
(and noting that $\sum_{i=0}^\ell \mu_i (r - \mu_i)$ is always even),
we obtain
\begin{eqnarray}
\Delta(\bldbeta^*)
& = &
\frac{1}{\det(V_r)}
\prod_{i=0}^\ell \det(X_i) \nonumber \\
& = &
\frac{1}{\det(V_r)}
\Bigl(
\prod_{i=0}^\ell
\alpha^{r_{i-1} \mu_i (r-1 - (\mu_i-1)/2)}
\Bigr) \nonumber \\
&& \makebox[10ex]{} \cdot
\Bigl(
\prod_{i=0}^\ell
\prod_{s=0}^{\mu_i-1} 
\prod_{t=s+1}^{r-1}
(\alpha^t - \alpha^s) 
\Bigr) \nonumber \\
\label{eq:altdethatA}
& = &
\frac{\alpha^{P-Q}}{\det(V_r)}
\prod_{i=0}^\ell
\prod_{s=0}^{\mu_i-1} 
\prod_{t=s+1}^{r-1}
(\alpha^t - \alpha^s) \; ,
\end{eqnarray}
where
\begin{eqnarray*}
P & = &
(r{-}1) \sum_{i=0}^\ell r_{i-1} \mu_i \\
& = &
(r{-}1) \sum_{0 \le k < i \le \ell} \mu_k \mu_i \\
& = &
\frac{r{-}1}{2} \Bigl( r^2 - \sum_{i=0}^\ell \mu_i^2 \Bigr)
\end{eqnarray*}
and
\[
Q =
\frac{1}{2} \sum_{i=0}^\ell r_{i-1} \mu_i (\mu_i{-}1) \; .
\]

Next, we express $\Delta(\bldbeta^*)$
using~(\ref{eq:determinant}).
Let the integer $N$ be defined by
\[
N = 
\sum_{0 \le i < k \le \ell}
\sum_{s=0}^{\mu_i-1}
\sum_{t=0}^{\mu_k-1}
(s + t + 1) \; .
\]
This integer can also be written as
\begin{eqnarray*}
N & = &
\sum_{0 \le i < k \le \ell}
\left(
  \frac{\mu_k \mu_i (\mu_i{-}1)}{2}
  + \frac{\mu_i \mu_k (\mu_k{-}1)}{2}
  + \mu_i \mu_k
\right) \\
& = &
\frac{1}{2}
\sum_{0 \le i < k \le \ell}
\mu_i \mu_k (\mu_i + \mu_k) \\
& = &
\frac{1}{2}
\sum_{i=0}^\ell \mu_i^2 (r - \mu_i) \; .
\end{eqnarray*}

  From~(\ref{eq:determinant}) we get
\begin{eqnarray*}
\lefteqn{
\Delta(\bldbeta^*)
} \makebox[0ex]{} \\
& = &
\kappa(\alpha) \cdot
\prod_{0 \le i < k \le \ell}
\prod_{s=0}^{\mu_i-1}
\prod_{t=0}^{\mu_k-1}
(\alpha^{r_k + s} - \alpha^{r_i + t}) \\
& = &
\kappa(\alpha) \cdot \alpha^N \cdot
\prod_{0 \le i < k \le \ell}
\prod_{s=0}^{\mu_i-1}
\prod_{t=0}^{\mu_k-1}
(\alpha^{r_k -t - 1} - \alpha^{r_i - s - 1}) \\
& = &
\kappa(\alpha) \cdot \alpha^N \cdot
\prod_{0 \le i < k \le \ell}
\prod_{s=0}^{\mu_i-1}
\prod_{t=0}^{\mu_k-1}
(\alpha^{r_{k-1} + t} - \alpha^{r_{i-1} + s}) \\
& = &
\kappa(\alpha) \cdot \alpha^N \cdot
\det(V_r) \\
&&
\makebox[10ex]{} {} \cdot 
\Bigl(
\prod_{i=0}^\ell
\prod_{s=0}^{\mu_i-1}
\prod_{t=s+1}^{\mu_i-1}
(\alpha^{r_{i-1} + t} - \alpha^{r_{i-1} + s}) \Bigr)^{-1} \\
& = &
\kappa(\alpha) \cdot \alpha^N \cdot
\det(V_r) \\
&&
\makebox[10ex]{} {} \cdot 
\left(
\prod_{i=0}^\ell
\Bigl(
\alpha^{r_{i-1} \mu_i(\mu_i-1)/2}
\det(V_{\mu_i}) \Bigr) \right)^{-1} \\
& = &
\kappa(\alpha) \cdot \alpha^{N-Q} \cdot
\det(V_r)
\Bigl( \prod_{i=0}^\ell \det(V_{\mu_i}) \Bigr)^{-1} \; .
\end{eqnarray*}
The last expression should be equal to~(\ref{eq:altdethatA}); so,
\begin{eqnarray*}
\lefteqn{
\kappa(\alpha)
} \makebox[0ex]{} \\
& = &
\frac{\Delta(\bldbeta^*) \cdot \alpha^{Q-N}}{\det(V_r)}
\prod_{i=0}^\ell \det(V_{\mu_i}) \\
& = &
\frac{\alpha^{P-Q+Q-N}}{(\det(V_r))^2}
\Bigl( \prod_{i=0}^\ell \det(V_{\mu_i}) \Bigr)
\prod_{i=0}^\ell
\prod_{s=0}^{\mu_i-1} 
\prod_{t=s+1}^{r-1}
(\alpha^t - \alpha^s) \\
& = &
\frac{\alpha^{P-N}}{(\det(V_r))^2}
\prod_{i=0}^\ell
\left(
(\det(V_{\mu_i}))^2
\prod_{s=0}^{\mu_i-1} 
\prod_{t=\mu_i}^{r-1}
(\alpha^t - \alpha^s)
\right)
\; ,
\end{eqnarray*}
where
\begin{eqnarray*}
P{-}N 
& = &
\frac{1}{2}
\left(
r^2 (r{-}1) -
(2r{-}1) \Bigl( \sum_{i=0}^\ell \mu_i^2 \Bigr) +
\sum_{i=0}^\ell \mu_i^3
\right) \\
& = &
\frac{1}{2}
\sum_{i=0}^\ell
\mu_i
\Bigl(
r (r{-}1) -
(2r{-}1) \mu_i + \mu_i^2
\Bigr) \\
& = &
\sum_{i=0}^\ell
\mu_i
{\tau_i \choose 2} \; .
\end{eqnarray*}

\end{document}